\documentclass[a4paper,11pt]{article}
\usepackage{graphicx}
\usepackage{url}

\title
{
Considerations on Classical and Quantum Bits
\thanks{The authors are undergraduate students of Computer Science at 
the Catholic University of Petr\'opolis. They are also members of Grupo de
F\'{\i}sica Te\'orica Jos\'e Leite Lopes.}
}
\author
{
F.L.~Marquezino and R.R.~Mello~J\'unior\\
\\
CBPF - Centro Brasileiro de Pesquisas F\'{\i}sicas \\
CCP - Coordena\c{c}\~ao de Campos e Part\'{\i}culas \\
Av. Dr. Xavier Sigaud, 150\\
22.290-180 Rio de Janeiro (RJ) Brazil\\
(CNPq Fellows/PIBIC)\\
{\it franklin@serraon.com.br, rui.rodrigues@inf.ucp.br}
}

\date{April, 2004}

\begin{document} 

\maketitle

\begin{abstract}
This article is a short review on the concept of information. We show the strong relation between Information Theory and Physics, beginning by the concept of bit and its representation with classical physical systems, and then going to the concept of quantum bit (the so-called ``qubit'') and exposing some differences and similarities.%
This paper is intended to be read by non-specialists and undergraduate students of Computer Science, Mathematics and Physics, with knowledge of Linear Algebra and Quantum Mechanics.
\end{abstract}

\section{Introduction}

Physics is an important subject in the study of information processing. It could not be different, since
information is always represented by a physical system. When we write, the information is encoded in ink particles
over a paper surface. When we think or memorize something, our neurons are storing and processing information. Morse code uses a physical system, such as light or sound waves to encode and transfer messages. As Rolf Laudauer said, ``\textit{information is physical}''. At least for the purposes of our study, this statement is very adequate.

Every day, we use classical systems to store or read information. This is part of human life since the very beginning of history. But, what happens if we use quantum systems instead of classical ones? This is an interesting subject in the intersection of Physics, Computer Science and Mathematics. 

In this article, we show how information is represented, both in quantum and classical systems. The plan of our work is as follows: in Section~\ref{sec:IP} we argue about the physical character of information. In Section~\ref{sec:Bit} we show the classical point of view of information, i.e., according to Newtonian Mechanics. In Section~\ref{sec:Qubit}, the point of view of Quantum Mechanics will be shown.

We also suggest some introductory references that explain most of the concepts discussed here~\cite{Portugal03,Franklin,Maser}.

The main goal of this paper is to review some mathematical and physical aspects of classical information and compare them with its quantum counterpart. 

\section{Information is physical}\label{sec:IP}

In its very beginning, Computer Science could be considered a branch of Mathematics, exclusively. However, since a few decades some scientists have been giving special attention to the correlation between Computer Science and Physics. 

One of the first physical aspects that we can raise in classical computation is thermodynamics. How much energy is spent when the computer does a certain calculation, and how much heat is dissipated? Is it possible to create a computer that does not spend any energy at all? To answer these questions we will begin by examining Landauer's principle.

According to Landauer's principle, when a computer erases a bit, the amount of energy dissipated is at least $k_B T \; \mbox{ln} \; 2$, where $k_B$ is Boltzmann's constant and $T$ is the temperature of the environment. The entropy of the environment increases at least $k_B \; \mbox{ln} \; 2$. This means that any irreversible operation performed by a computer dissipates heat and spends energy. For instance, the AND logical operation\footnote{If the reader is not familiar with the concept of logical gate, we recommend the reading of \cite{operat}.} is irreversible, because given an output we cannot necessarily know the inputs. If the output is $0$, the inputs could be $00$, $01$ or $10$. This operation erases information from the input, so it dissipates energy, according to Landauer's principle.

If one could create a computer using only reversible operations, this computer would not spend any energy. That would be a great achievement, given the fact that our modern society spends more and more in energy, and computers are responsible for great part of the problem. Charles Bennett, in 1973, proved that building a reversible computer is possible~\cite{Bennett73}. The next step would be finding universal reversible gates, i.e., a gate or a small set of gates that allows the construction of circuits to calculate any computable function. E. Fredkin and T. Toffoli proved the existence of such gate in 1982~\cite{Toffoli82}. The Toffoli gate is equivalent to the traditional NAND operation (which is universal in classical computation) and works as follows:

\begin{equation}
\mbox{Toffoli}(A,B,C)=(A,B,C \oplus A \wedge B),
\end{equation}
where $\oplus$ is sum modulus 2 and $\wedge$ in the logical AND. 

In priciple we could build a reversible computer by simply replacing NAND gates by Toffoli gates. That is not so simple to implement, though. Besides, one could question whether this gate is actually non-dissipative, since we generate a lot of junk bits that will need to be erased sometime. Bennett solved this problem by observing that we could perform the entire computation, print the answer (which is a reversible operation in Classical Mechanics) and then run the computer backwards, returning to the initial state. So, we do not need to erase the extra bits.

Another interesting subject that leads us to the intersection between Computer Science and Physics is the Maxwell's demon. In 1871, J.C. Maxwell proposed a theoretic machine, operated by a little ``demon'' that could  violate the second law of thermodynamics~\cite{Maxwell}. 

The machine would have two partitions, separated by a little door controlled by this demon. The {\it modus operandi} of this demon would be quite interesting. It would watch the movement of each molecule, opening the door whenever they approach, allowing fast ones to pass from the left to right partition and slow ones to pass from right to left partition. By doing that, heat would flow from a cold place to a hot one at no cost. The solution for this apparently paradox resides in the fact that the demon must store information about the movement of the particles. Since the demon's memory is finite, it will have to erase information in a moment, dissipating energy and then increasing the entropy of the system.

The topics pointed out in this section show how close Computer Science and Physics are. In the next sections we will show how information is represented by Classical Mechanics, and what happens if we use Quantum Mechanics instead.

\section{On classical bits}\label{sec:Bit}

A classical computer performs logical and arithmetical operations with a certain (finite) alphabet\footnote{The Turing machine was proposed by Alan Turing in 1936 and became very important for the understanding of what computers can do~\cite{Turing}. It is composed by a program, a finite state control, a tape and a read/write tape head.}. 
Each one of the symbols that compose this alphabet must be represented by a specific state of a classical system.  

Since we are used to perform calculations with decimal numbers it is very natural to think that the computer's alphabet should be composed by ten different symbols. However, it would be very expensive and complex to build a computer with this characteristic. Instead, computers work with 2-state systems, the so-called \textit{bits}, and represent binary numbers.

The concept of \textit{bit} was anticipated by Leo Szilard~\cite{Szi} while analyzing the Maxwell's demon paradox. However, the word \textit{bit} (binary digit) was first introduced by Tukey. The \textit{bit} is the fundamental concept in Information Theory, and is the smallest information that can be handled by a classical computer.
Every information stored in the computer is either a bit or a sequence of bits.

If we join $n$ bits, we can represent $2^n$ different characters. But, how many bits are necessary to represent all the characters in the English alphabet, plus the numbers and some special characters? If we use 8 bits, we can represent 256 characters, which is enough! To these 8 bits we give the name \textit{byte}\footnote{Some authors say that the group of 8 bits are special because of the 80x86 processor. This processor used 8 bits to give  memory addresses, i.e., it had 256 different addresses in memory.}. Another interesting unit is the nibble, which is formed by 4 bits. With one nibble we can represent all the hexadecimal numbers ($2^4 = 16$). Since the hexadecimal base is largely used in assembly languages and low-level computing, some computer scientists work with nibbles quite often.

The byte is a very small unit, so we normally use some of its multiples. The kilobyte (KB) corresponds to 1024 bytes, i.e., 8192 bits. One could think that 1KB should be 1000 bytes, but as we are dealing with binary numbers, the power of 2 which is closer to 1000 is actually $2^{10} = 1024$.

There are also some other useful units: megabyte (MB), which corresponds to 1024 KB, gigabyte (GB), equals to 1024 MB, terabyte (TB), equivalent to 1024 GB and petabyte, which corresponds to 1024 TB.

At this point, the idea of Shannon entropy should be introduced~\cite{ShannonEntr}.
Shannon entropy is an important concept of Information Theory, which quantifies the uncertainty about a physical system. We can also look at Shannon entropy in a different point of view, as a function that measures the amount of information we obtain, on average, when we learn the state of a physical system. 

We define Shannon entropy as a function of a probability distribution, $p_1, p_2, \ldots, p_n$:

\begin{equation}
H(p_1, p_2, \ldots, p_n) \equiv -\sum_{x}{p_x \log{p_x}}
\end{equation}
where $0 \log{0} \equiv 0$, in the context of distributions or generalized functions. Note that $\lim_{x \rightarrow 0}{(x \log{x})}=0$.

This function will be explained in this paper through an exercise, which can also be found in \cite{Chuang00}. This is an intuitive justification for the function we defined above. 

Suppose we want to measure the amount of information associated to an event E, which occurs in a probabilistic experiment. We will use a function $I(E)$, which fits the following requirements:

\begin{enumerate}
\item $I(E)$ is a function only of the event $E$, so we may write $I=I(p)$, where $p$ is the probability of the event $E$;
\item $I$ is a smooth function of probability;
\item $I(pq)=I(p)+I(q)$ when $p,q>0$, i.e., the information obtained when two independent events occur with probabilities $p$ and $q$ is the sum of the information obtained by each event alone.
\end{enumerate}

We want to show that $I=k \log{p}$, for some constant $k$. From the third condition of the problem, 

\begin{equation}
I(pq)=I(p)+I(q),
\end{equation} 
we can let $q=1$, verifying that $I(1)=0$. Now, we can differentiate both sides of the above equation with respect to $p$.

\begin{eqnarray}
\frac{\partial I(pq)}{\partial p}=\frac{dI(p)}{dp} + \frac{dI(q)}{dp}\\
\frac{d I(pq)}{d (pq)} \cdot \frac{\partial pq}{\partial p}=I'(p)\\
I'(pq) \cdot q = I'(p).
\end{eqnarray}

When $p=1$ we can easily note that
\begin{equation}
I'(q) \cdot q = I'(1).
\end{equation}

Based on the second condition of the problem, we know that $I'(p)$ is well defined when $p=1$, so $I'(1)=k$, $k$ constant.

\begin{eqnarray}
I'(q)=\frac{k}{q}\\
I(q)=\int{\frac{k}{q}dq}\\
I(q)=k \log{q}.
\end{eqnarray}

The function $I(p)$ appeared naturally and satisfies the three conditions specified by the problem. However, the function $I(p)$ represents the amount of information gained by one event with probability $p$. We are interested in a function that gives us the mean information, that is, the entropy.

\begin{eqnarray}
H = <I> = \frac{\sum_x{p_x(k \log{p_x})}}{\sum_x{p_x}}\\
H= <I> = k \sum_x{p_x \log{p_x}}
\end{eqnarray}
where $k = -1$, and we have the Shannon entropy's formula:

\begin{equation}
H = - \sum_x{p_x \log{p_x}}.\label{eq:entr}
\end{equation}

If we apply (\ref{eq:entr}) specifically to the case where we have a binary random variable (which is very common in Computer Science), this entropy receive the name {\it binary entropy}:

\begin{equation}
H_{bin}(p)=-p \log{p} - (1-p) \log{(1-p)}
\end{equation}
where $p$ is the probability for the variable to have the value $v$, and $(1-p)$ is the probability for the variable to assume the value $\neg v$.

Information Theory studies the amount of information contained in a certain message, and its transmission through a channel. Shannon's Information Theory was responsible for giving a precise and mathematical definition for information.

Written languages can be analyzed with the help of Information Theory~\cite{Bruce}. For a given language, we can define the \textit{rate of the language} as

\begin{equation}
r=\frac{H(M)}{N},
\end{equation}
where $H(M)$ is the Shannon entropy of a particular message and $N$ is the lenght of this message. 

In English texts, $r$ normally varies from $1.0$ to $1.5$ bit per letter. Cover found $r=1.3$ bits/letter in~\cite{Thomas}. Assuming that, in a certain language composed by $L$ characters, the probability of occurence of each letter is equal, one can easily found the amount of information contained in each character. 

\begin{equation}
R=\log{L},\label{eq:abs-redundancy}
\end{equation}
where $R$, the maximum entropy of each character in a language, is called \textit{absolute rate}. 

The English alphabet is composed by 26 letters, so its absolute rate is $\log{26} \approx 4.7$ bits/letter. The absolute rate is normally higher that the rate of the language. Hence, we can define the \textit{redundancy} of a language as

\begin{equation}
D=R-r.
\end{equation}

In the English language, if we consider $r=1.3$ according to~\cite{Thomas}, and if we apply eq.~(\ref{eq:abs-redundancy}) to find $R$, we find out that the redundancy is $3.4$ bits/letter.

We cannot forget that we deal with information every day. In this exact moment, you are dealing with the information contained in this paper. So, it is natural to ask how much information our senses can deal with. Studies have shown that vision can receive $2.8 \cdot 10^8$ bits per second, while audition can deal with $3 \cdot 10^4$ bits per second. Our memory can store and organize information for a long time. The storage capacity of the human brain varies from $10^{11}$ to $10^{12}$ bits. Just as a comparison we can mention that the knowledge of a foreign language requires about $4 \cdot 10^6$ bits~\cite{Kupf}.

\section{Introducing qubits}\label{sec:Qubit}

The quantum bit is often called ``qubit''. The key idea is that a quantum system will be used to store and
handle data. When we use a classical system, such as a capacitor or a transistor, the properties of
Classical Mechanics are still observed. On the other hand, if we use a quantum system to process information,
we can take advantage of the quantum-mechanical properties.

Quantum Mechanics has a probabilistic character. While a classical system can be in one, and only one state, a quantum system can be in a state of superposition, as if it was in different states simultaneously, each one associated to a probability. Mathematically, we will express this $N$-state quantum system as the linear combination,

\begin{equation}
| \psi \rangle = \sum_{i=0}^{N-1}{a_i | i \rangle }
\end{equation}
where $a_i$ are complex numbers called amplitudes. We know, from the Quantum Mechanics postulates, that $\| a_k \|^2$ is the probability of obtaining $| k \rangle$ when measuring the state $| \psi \rangle$. Then,

\begin{equation}\label{eq:norm}
\sum_{i=0}^{N-1}{ \| a_i \|^2 } = 1.
\end{equation}

In Quantum Computation we normally work with 2-state systems (otherwise we would not be referring to qu\textbf{bits}, but qutrits, qu-nits or something similar). So, the quantum bits can assume any value in the form:

\begin{equation}\label{eq:qubit}
| \psi \rangle = \alpha | 0 \rangle + \beta | 1 \rangle
\end{equation}
with $\alpha$, $\beta$ complex numbers, and $\| \alpha \|^2 + \| \beta \|^2=1$. It is important to stress that the amplitudes are not simple probabilities. The state $\frac{1}{\sqrt{2}}(| 0 \rangle + | 1 \rangle)$ is different from $\frac{1}{\sqrt{2}}(| 0 \rangle - | 1 \rangle)$, for instance. In this case we say that the two states are different by a relative phase. However, the states $| \psi \rangle$ and $e^{i\theta} | \psi \rangle$ (where $\theta$ is a real number) are considered equal, because they differ only by a global phase. The global phase factor does not have influence in the measurement of the state.

Superposition is quite interesting because while classical bits can assume only one value, its quantum counterpart can assume a superposition of states. A single qubit can value both $0$ and $1$ simultaneously. Similarly a $n$-qubit register can value, simultaneously, all the values from $0$ to $2^n -1$. Consequently, we can do the same calculation on different values at the same time, simply by performing an operation on a quantum system. 

Now, returning to the mathematical study of the quantum system, we can observe that a single qubit can be represented in a two-dimensional complex vector space. Of course, that does not help us so much in terms of geometric visualization. However, note that we may rewrite eq.~(\ref{eq:qubit}):

\begin{equation}
| \psi \rangle = e^{i \gamma} \biggl( \cos{\frac{\theta}{2}} | 0 \rangle + e^{i \varphi} \sin{\frac{\theta}{2} | 1 \rangle} \biggl)
\end{equation}
where $\theta$, $\varphi$ and $\gamma$ are real numbers. The global factor $e^{i \gamma}$ can be ignored, since it has no observable effects.

\begin{equation}
| \psi \rangle = \cos{\frac{\theta}{2}} + e^{i \varphi} \sin{\frac{\theta}{2}} | 1 \rangle.
\end{equation}

Now, we can represent a qubit in a three-dimensional real vector space. According to eq.~(\ref{eq:norm}), the qubit norm must be equal to 1, so the numbers $\theta$ and $\varphi$ will define a sphere: the so-called \textit{Bloch sphere}.

\begin{figure}
\centering
\includegraphics[width=170pt]{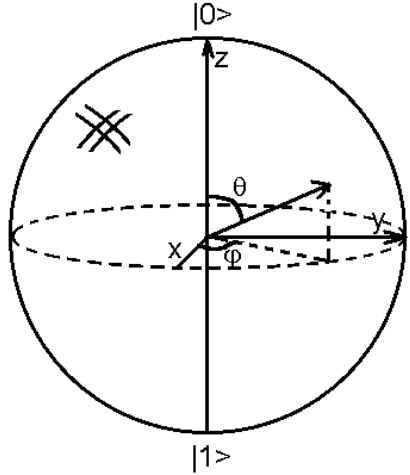}
\caption{Bloch sphere.}
\label{fig:bloch}
\end{figure}

As we can see, there are infinite points on the Bloch sphere. Nevertheless, it is important to emphasize that all we can learn from a measurement is 0 or 1, but not the values of $\theta$ or $\varphi$. Moreover, after performing a measurement the state will be irreversibly collapsed (projected) to either $| 0 \rangle$ or $| 1 \rangle$. Should it be different, we could write an entire encyclopedia in one qubit, by taking advantage of the infinite solutions of (\ref{eq:norm}).

If we wished to represent a composite physical system (which could be a quantum register, for instance), we would use an operation called tensor product, represented by the symbol $\otimes$. The state of a quantum register $| \phi \rangle$ composed by the qubits $| \psi_i \rangle$, where $i$ varies from 1 to $n$ is

\begin{equation}
| \phi \rangle = | \psi_1 \rangle \otimes | \psi_2 \rangle \otimes \ldots \otimes | \psi_n \rangle.
\end{equation}

We recommend that the reader refers to~\cite{Chuang00} to get more information on this postulate.

\subsection{The no-cloning theorem}

There is a remarkable difference between classical and quantum states, which is the impossibility of the latter to be perfectly cloned when it is not known \textit{a priori}. This can be proved by the \textit{no-cloning} theorem, published by W.K. Wooters and W.H. Zurek, in 1982~\cite{Wooters82}. Here, we will prove that a generic quantum state cannot be cloned. The authors recommend the reading of the article cited before for a more complete comprehension.

Let us suppose we wish to create a machine that receives two qubits as inputs, called qubit A and qubit B. Qubit A will receive an unknown quantum state, $| \psi \rangle$, and qubit B a pure standard state, $| s \rangle$ (such as a blank sheet of paper in a copy machine). We wish to copy the state $| \psi \rangle$ to qubit B. The initial state of the machine is

\begin{equation}\label{eq:unitary}
| \psi \rangle \otimes | s \rangle.
\end{equation}

If the copy was possible, there were an unitary operator $U$ such that $U(| \psi \rangle \otimes | s \rangle) = | \psi \rangle \otimes | \psi \rangle$. However, we wish our machine to be able to copy different states. So, the operator U must be such that $U(| \phi \rangle \otimes | s \rangle) = | \phi \rangle \otimes | \phi \rangle$. The inner product between these two equations is

\begin{equation}
\langle \psi | \phi \rangle = ( \langle \psi | \phi \rangle )^2.
\end{equation}

It is easy to realize that the solutions for this equation are $\langle \psi | \phi \rangle = 1$ and $\langle \psi | \phi \rangle = 0$, i.e., when $| \phi \rangle = | \psi \rangle$, or when $| \phi \rangle \perp | \psi \rangle$. The first solution is useless, so we proved that a perfect cloning machine is only able to clone orthogonal states.

The non-cloning theorem leads us to a very interesting application of Quantum Mechanics: a provable secure protocol for key distribution that can be used together with Vernam's cipher to provide an absolutely reliable cryptography. The reader can refer to~\cite{Franklin} for a short introduction to this subject.

\subsection{Von Neumann entropy}

Up to this point, we have been using the vector language to express Quantum Mechanics. From now on, it will be interesting to introduce another formalism: the density operator (also called ``density matrix''). This is absolutely equivalent to the language of state vectors, but it will make the calculations much easier in this case. Besides, the density operator is an excellent way to express quantum systems whose state is not completely known. If we have a quantum system with probability $p_i$ to be in the state $| \psi_i \rangle$, then we call $\{ p_i, | \psi_i \rangle \}$ an ensemble of pure states. We define the density matrix for this system as

\begin{equation}
\rho = \sum_i{p_i}|\psi_i \rangle \langle \psi_i |.
\end{equation}

Von Neumann entropy is very similar to the Shannon entropy. It measures the uncertainty associated with a quantum state. The quantum state $\rho$ has its Von Neumann entropy given by the formula

\begin{equation}
S(\rho)=-tr(\rho \log{\rho}).
\end{equation}

Let $\lambda_i$ be the eigenvalues of $\rho$. It is not very difficult to realize that the Von Neumann entropy can be rewritten as

\begin{equation}
S(\rho)=-\sum_x{\lambda_x \log{\lambda_x}}.
\end{equation}

Another important concept is the relative entropy. We can define the relative entropy of $\rho$ to $\sigma$ as

\begin{equation}
S(\rho || \sigma) = tr(\rho \log{\rho}) - tr(\rho \log{\sigma})
\end{equation}
where $\rho$ and $\sigma$ are density operators.

According to Klein's inequality, the quantum relative entropy is never negative:

\begin{equation}
S(\rho || \sigma) \geq 0
\end{equation}
with equality holding if and only if $\rho = \sigma$. The proof for this theorem is not relevant here, but it can be found in~\cite[page 511]{Chuang00}.

\subsection{Further comments on Quantum Information Theory}

The Quantum Information Theory is concerned with the information exchange between two or more parties, when a quantum mechanical channel is used to achieve this objective. Naturally, the purpose of this paper is not to give a deep comprehension of this subject. Quantum Information Theory, as well as its classical counterpart, is a vast area of knowledge, which would require much more than just few pages to be fully explained. Instead, we give some basic elements, allowing the reader, independently of his area of knowledge, to have a better comprehension of Quantum Computation and Quantum Information Processing. 

Quantum systems have a collection of astonishing properties. Some of them could, at least in principle, be used in Computer Science, allowing the production of new technology. One of these amazing properties we have already mentioned: it is the superposition. If in the future mankind learn how to control a large number of qubits in a state of superposition for enough time, we will have the computer of our dreams. It would be a great step for science.

Another important property is the entanglement~\cite{Chuang00,Preskill}. Some states are so strongly connected that one cannot be written disregarding the other. In other words, they cannot be written separately, as a tensor product. This property brings interesting consequences. Imagine that Alice prepares the state below\footnote{This is one of the so-called Bell states.} in her laboratory, in Brazil:

\begin{equation}
|\beta_{00}\rangle = \frac{|0\rangle_a |0\rangle_b + |1\rangle_a |1\rangle_b}{\sqrt{2}}.
\end{equation}

After that, Alice keeps qubit $a$ and gives qubit $b$ to Bob, who will take it to another laboratory, let us say, in Australia. Now, we know from the third postulate of Quantum Mechanics that if any of them measure the state, it will collapse either to $|0\rangle_a |0\rangle_b$ or to $|1\rangle_a |1\rangle_b$. So, the state of the qubit in Australia can be modified by a measurement done in Brazil and vice-versa!

Reference~\cite{Chuang00} is strongly recommended as a starting point, for those who want to study this topic more deeply.

\section{Concluding remarks}

In this paper, we have shown some of the main aspects of information. Information Theory normally considers that all information must have a physical representation. But, Nature is much more than the classical world, that we see every day. If we remember that the amazing quantum world can also represent information, we discover astonishing properties, leading us to a new field of study. Here, we briefly introduced this subject to students and researchers from different areas of knowledge.

In Computer Science, we normally wish to represent some information, manipulate it in order to perform some calculation and, finally, measure it, obtaining the result. We began by showing how information is represented, in classical systems and in quantum systems. In a forthcoming work~\cite{operat}, we show how information can be manipulated in each case.

Both classical and quantum information have similarities and differences, that were quickly exposed in this article. The technological differences are still enormous. While the technology to produce classical computers are highly developed, the experiments involving quantum computers are not so simple and have a slow progress. However, as we saw in this article, the properties of quantum information are so interesting that the development of quantum computers in the future can become one of the greatest achievements of our history.

\section*{Acknowledgements}

The authors thank Prof. J.A. Helay\"el-Neto (CBPF) and Dr. J.L. Acebal (PUC-Minas) for reading the manuscripts, and for providing helpful discussions. We thank the Group of Quantum Computation at LNCC, in particular Drs. R.~Portugal and F.~Haas, for the courses and stimulating discussions. We also thank the brazilian institution CNPq and the PIBIC program, for the financial support.

\end{document}